\documentclass{sig-alternate-05-2015}
\usepackage{graphicx}
\usepackage{amssymb}
\usepackage{algorithm}
\usepackage{algpseudocode}
\usepackage{textcomp}
\usepackage{listings}
\usepackage[usenames,dvipsnames]{xcolor}
\usepackage{subcaption}
\usepackage{cite}
\usepackage{hyperref}
\usepackage{lipsum}
\usepackage[normalem]{ulem}
\usepackage{listings}
\usepackage{xspace}
\usepackage{color}
\usepackage{wrapfig}
\usepackage[textsize=tiny]{todonotes}
\usepackage{cleveref}
\usepackage{tabu}
\usepackage{paralist}

\newcommand{\tuple}[1]{\left<\;{#1}\;\right>}

\newcommand{\tinysection}[1]{\smallskip\noindent \textbf{#1.}$\,$}

\newcommand{\inlineitem}[1]{\noindent -- \emph{{#1}:}}

\newcommand{\hidecomment}[1]{}

\newcommand{\sysname}{Vizier\xspace}
\newcommand{\langname}{\textsc{VizUAL}\xspace}

\setcopyright{acmcopyright}
\toappear{}

\title{The Exception that Improves the Rule\titlenote{The authors are listed in alphabetical order.}}

\numberofauthors{1}
\author{
\alignauthor
Juliana Freire$^n$, Boris Glavic$^i$, Oliver Kennedy$^b$, Heiko Mueller$^n$\\
       \affaddr{$n$: New York University;  \{juliana.freire, heiko.mueller\}@nyu.edu}\\
       \affaddr{$i$: Illinois Institute of Technology; bglavic@iit.edu}\\
       \affaddr{$b$: University at Buffalo; okennedy@buffalo.edu}
}

\begin{document}

\maketitle
\begin{abstract}
The database community has developed numerous tools and techniques for data curation and exploration, from declarative languages, to specialized techniques for data repair, and more.  
Yet, there is currently  no consensus on how to best expose these powerful tools to an analyst in a simple, intuitive, and above all, flexible way.
Thus, analysts continue to rely on tools such as spreadsheets, imperative languages, and notebook style programming environments like Jupyter for data curation.
In this work, we explore the 
integration of spreadsheets, notebooks, and relational databases.
We focus on a key advantage that both spreadsheets and imperative notebook environments have over classical relational databases: ease of exception.  
By relying on set-at-a-time operations, relational databases sacrifice the ability to easily define singleton operations, exceptions to a normal data processing workflow that affect query processing for a fixed set of explicitly targeted records.  
In comparison, a spreadsheet user can easily change the formula for just one cell, while a notebook user can add an imperative operation to her notebook that alters an output ``view''.  
We believe that enabling such idiosyncratic manual transformations in a classical relational database is critical for curation, as curation operations that are easy to declare for individual values can often be extremely challenging to generalize.  
We explore the challenges of enabling singletons in relational databases, propose a hybrid spreadsheet/relational notebook environment for data curation, and present our vision of \sysname, a system that exposes data curation through such an interface.  
\end{abstract}


\section{Introduction}
\label{sec:introduction}

In spite of the availability of powerful automated curation, cleaning, and analysis tools, spreadsheets and notebook UIs (e.g., Jupyter/iPython) are still the predominant tools used by most data scientists for manipulating and visualizing virtually all but the largest datasets. Although their ubiquity is in part a matter of user familiarity~\cite{Chan1996119}, we argue that they also offer several compelling benefits, specifically for curation workloads.
Key among these is the simplicity with which users can define \emph{exceptions} to bulk set-at-a-time operations in both spreadsheets and notebooks.  
In this paper, we examine these two interface models and explore how lessons from both can be incorporated into relational database interfaces.
We present a new \emph{data curation} user interface and a tool implementing this interface called \sysname.
\sysname will combine UI elements from both spreadsheets and notebooks and support functionality not commonly found in either spreadsheets or notebooks, including automated curation operators~\cite{Yang:2015:LOA:2824032.2824055}, deployment of workflows over large datasets~\cite{Kandel:2011:WIV:1978942.1979444}, declarative queries~\cite{AG16,Olston:2008:PLN:1376616.1376726}, and support for exploratory curation tasks~\cite{SV08}. 
This hybrid UI enables powerful relational queries, but remains flexible enough to permit easy data manipulation, summarization, and visualization.

\vspace{-.2cm}
\tinysection{Spreadsheets}
Spreadsheets are a ubiquitous data processing tool.  Their simplicity, generality, and adaptability make them ideal for ``playing'' with data through predominantly visual programming metaphors. 
Spreadsheets provide several important features that are useful during data curation:\\
\inlineitem{Convenient modification of values and computations} The user can update any cell's value or formula directly from the user interface. This enables manual curation operations like resolving missing values and 
correcting data  errors.\\
\inlineitem{Manual operations with inlined results} By using formulas in cells, a user-defined computation and its result are shown together with its input data.\\
\inlineitem{Visual mapping over data collections} 
Most spreadsheet systems enable the user to take a formula (computation) and map it to a range of cells through \textit{position-relative} references in cell formulas. 
For example, this can be done by copy/paste or by a fill operation. We refer to this mechanism as \textit{adapt\&apply}. This approach to bulk, set-at-a-time functionality is very useful in data curation: A fix to repair one piece of data (e.g., conversion between units) can be deployed over the whole dataset. 
Importantly, the formulas applied in this fashion are  independent of each other  creating an affordance for declaring exceptions to the bulk rule by allowing individual formulas to be modified.



Indeed, many curation applications require users to ``break the rules'' and apply one-off modifications or transformations to individual fields or records.  
For example, 
(1) hypothetical what-if scenarios require users to apply small ad-hoc updates to adjust the inputs under test; 
(2) repairs for data errors may be easy to define for individual cases, but far harder to define in a general case; 
(3) complex data transformations that need to be generalized would still be easier to define for individual test cases than in bulk.
By making it easy for users to break the rules, even if only temporarily, spreadsheets empower users to explore data, evaluate options, and better understand the effects of their curation efforts.  
Such exceptions, or \textit{singleton} operations, are not handled gracefully by existing relational DBMSes.
However spreadsheets also have several drawbacks compared to a DBMS:\\
\inlineitem{Non-Adaptive Computation over Collections} 
Spreadsheets 
allow a computation defined in one cell to be adapted to larger collections of cells.  However, the two dominant systems, i.e., Microsoft Excel and Google Sheets, do not support automatically extending formulas to new cells as data is added~\footnote{We note that Apple Numbers does exhibit this behavior.}.
This is in stark contrast to relational databases with their declarative, data-independent query languages. \\
\inlineitem{Collection operation intent is not explicit}
Adapt\&apply allows a computation to be mapped over a collection, but there is no visual cue that indicates that a set of cells are storing formulas which were mapped in this way. There is no abstraction in spreadsheets to represent higher-level bulk operations such as view queries in a database.\\
\inlineitem{No order among operations or workflow branch tracking} 
Spreadsheets use cell highlighting as a visual metaphor for dependency tracking.  
These visualizations are specific to single cells and do not lend themselves to tracking large curation workflows.
Furthermore, these visualizations are limited to tracing one dependency at a time, making tracking transitive dependencies cumbersome.\\
\inlineitem{Unintuitive results for adapt\&apply} 
As we will discuss further in Section~\ref{sec:language}, adapt\&apply functionality as implemented in many systems can lead to unexpected results.


\tinysection{Notebook-style UIs}
Systems like Jupyter expose an interactive, interpreted programming environment through a notebook-like interface where the user mixes documentation (text) with code. The output for code blocks is shown directly in the notebook --- a feature that is widely used to produce data visualizations.\\
\inlineitem{Inline documentation} Notebooks allow users to integrate comments, descriptive text, notes, and formatting details, making it easier for others to retrace their steps.\\
\inlineitem{Incremental development of complex workflows} Notebooks allow users to incrementally build curation workflows, one page at a time.  The (always linear) structure of the workflow is made explicit through the notebook interface.
\smallskip 

However, some operations that are efficiently supported in spreadsheets are harder to express in notebooks, and some disadvantages are shared among both paradigms:\\
\inlineitem{Small edits are cumbersome} Compared to spreadsheets, modifying individual data values requires users to write code.\\
\inlineitem{Linear workflows and no backtracking} Notebooks are inherently linear and do not allow users to backtrack or branch their development efforts.\\
\inlineitem{All-at-once processing} 
Both spreadsheets and notebooks operate over datasets in their entirety, something that is not feasible for large giga-/tera-byte files.

\tinysection{Visualizations}
Both spreadsheets and notebook UIs make it very easy for users to create  visualizations from data on the fly and show these visualization inline with the data. Also both paradigms allow these visualization to be tweaked and to be refreshed based on changes to their inputs. Spreadsheets in particular provide a very easy-to-use interface for selecting what data should be visualized.

\tinysection{Combined Spreadsheets and Notebooks} 
Spreadsheets permit intuitive visual interactions with data, while notebooks provide a clearer expression of the user's intent that can actually be reproduced.  We propose a hybrid UI that combines elements of both interfaces, augmenting them with capabilities common to relational data processing.  We discuss the challenges of developing such an integrated interface and how it facilitates data curation and exploration.  We also introduce our proposed system, called \sysname, which empowers users with spreadsheet-like flexibility for transforming, visualizing, and exploring relational data, while still retaining the expressiveness and workflow capabilities of a notebook.  At the heart of our approach is support for singleton operations in a relational setting, which in turn enables a bi-directional mapping between a spreadsheet-style graphical interface, and a notebook-style programmatic interface.

To enable singleton transformations within the framework of a classical relational database, we extend the notebook programming model with support for \textit{interactive views}.  
An interactive view begins its life as a classical view, presented to the user in tabular form.  
In contrast to a classical view however, an interactive view can be edited much like a spreadsheet. Such updates are not propagated back to the view's inputs, but are treated as updates to the view definition.
Users can modify fields, add new rows and columns, use a spreadsheet-style equation editor to define derived values, and more.  
As the user edits the view, the user's actions are seamlessly transformed into a program of relational(-ish) data transformation operators that derive the new, edited view.  
This program serves as a form of history, allowing the user to revisit and revise earlier edits, even out of order.  Furthermore, the program defines a workflow, highly specialized to a specific dataset.  
Even this is sufficient to provide classical benefits of workflow provenance such as auditability and explainability for derived data.  
Once an interactive view is developed for one dataset, it can more readily be adapted to new data or to react to changes in its inputs.  
Recasting the user's actions programmatically allows us to leverage existing work on algebraic equivalences~\cite{Liu:2009:SAD:1546683.1547431} and program rewriting~\cite{AG16} to first obtain multiple interpretations of sequences of user actions, and then to extrapolate more general expressions of the user's intent~\cite{Deutch:2016aa,Zloof:1975:QE:1499949.1500034}.  
An important challenge is controlling unexpected side effects arising from these edits.
Unlike a relational database where query semantics are explicit, visual interactions in a spreadsheet trigger many implicit behaviors.  
Existing spreadsheet software is carefully designed to ensure that these behaviors follow a consistent, intuitive pattern, and we ensure that \sysname retains this consistent behavior through a data model that links relational data with the spreadsheet interface's coordinate system.

\tinysection{Contributions} 
Our core contributions are as follows:
(1) We present \sysname, a hybrid relational notebook/spreadsheet exploration-based data curation framework and outline its capabilities,
(2) We discuss the challenges of mapping actions back and forth between the different components,
(3) We define a data model for \sysname that allows us to precisely characterize the side-effects of a user's actions on a spreadsheet.
(4) We apply this model through a case study on existing spreadsheet software, and show how user actions in spreadsheet software follow a specific heuristic that we believe captures the principle of least surprise~\cite{saltzer2009principles}.
(5) We outline future research directions, including a readability-optimizer and generalization of singleton-based workflows.


\vspace{-.3cm}
\section{Interface Design}
\label{sec:interface}

\begin{figure}
  \centering
  \includegraphics[width=0.75\columnwidth]{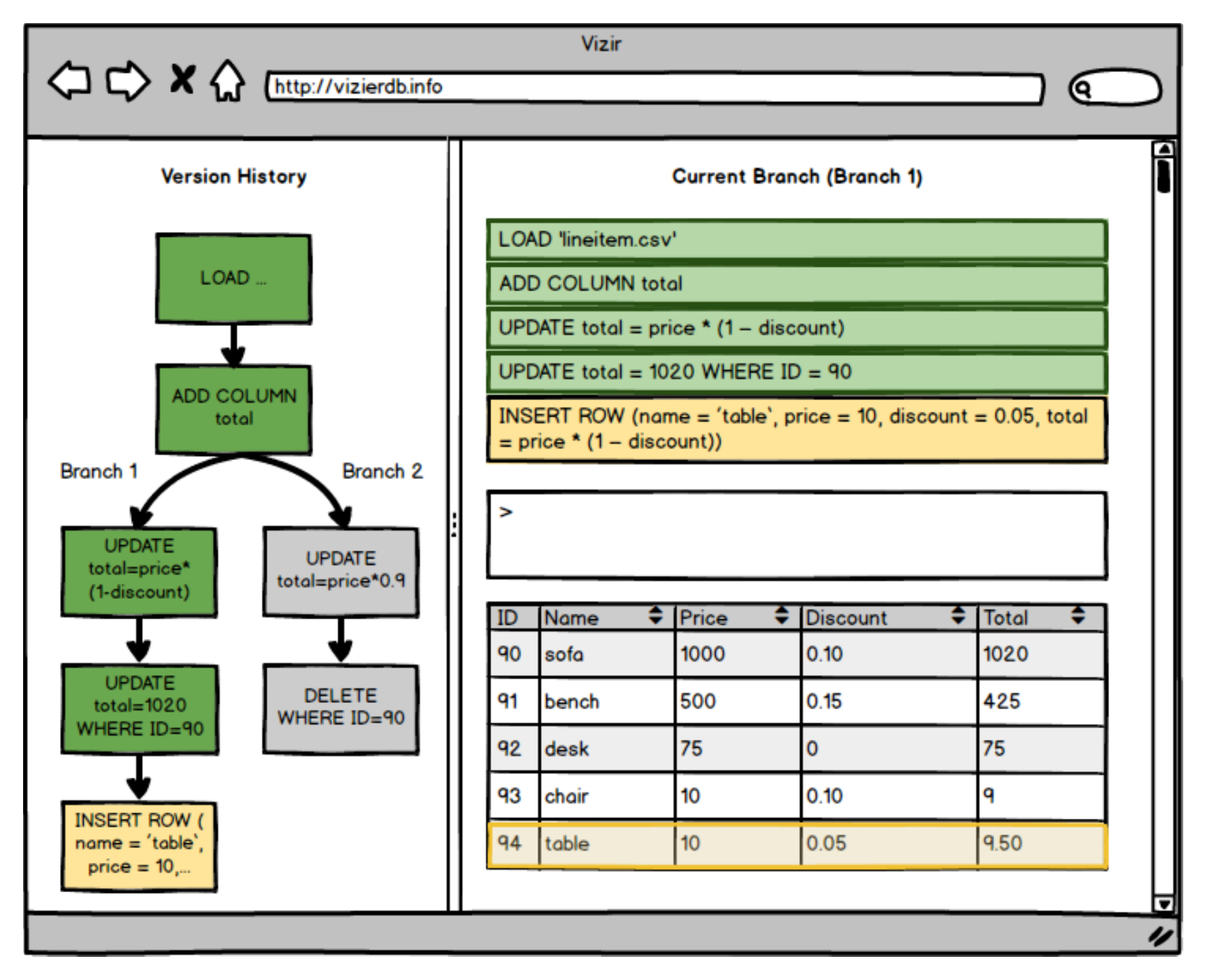}
  \caption{An example of \sysname's UI}
    \label{fig:hybridinterface}
    \vspace*{-3mm}
\end{figure}

Figure~\ref{fig:hybridinterface} illustrates the interface for \sysname, our proposed tool for data curation and exploration.  This interface combines elements of both notebooks and spreadsheets.  

\subsection{The Notebook UI}

Notebook interfaces like Jupyter's use an analogy of pages in a notebook that consist of a block of code and an output for the block, e.g., a table, visualization, or documentation.  Blocks are part of a continuous program, allowing a user to quickly probe intermediate states by creating new visualizations or views of the data.

Each page in a \sysname notebook includes a block of SQL DML/DDL code that imperatively manipulates a relation that is displayed as a table or visualization.  Pages are evaluated in sequential order.  Code defining later pages may reference preceding pages as if they were views, and edits to a page may result in cascading changes to pages that depend on it.  
We refer to this SQL-based language as the \sysname user action language (\langname).  
In spite of its imperative flavor, operators in \langname form a monad that can be compiled down to a generalized form of relational algebra~\cite{AG16}.

\begin{figure}
\begin{lstlisting}[morekeywords={LOAD,ROW}]
LOAD 'lineitem.csv';
ADD COLUMN total;
UPDATE total = price * (1 - discount);
UPDATE total = 1020 WHERE ID = 90;
INSERT ROW ( name = 'table', price = 10, 
             discount = 0.05, total = 9.5 );
\end{lstlisting}
\caption{An example \langname script}
\label{fig:program}
\vspace*{-5mm}
\end{figure}

Figure~\ref{fig:program} shows an example \langname script that loads a CSV file, extends it with a new column named \texttt{total}, defines a value for the column (derived from the remaining attributes), and applies two minor \emph{singleton} edits to the result (a single value update and a row pasted into the result).  
The script defines a sequence of declarative transformations on the data imported by the \texttt{LOAD} operation in the first line.  
The entire script can be rewritten into a SQL query:
\begin{lstlisting}[morekeywords={LOAD}]
SELECT *, CASE WHEN ID = 90 THEN 1020 
               ELSE price*(1-discount)
          END AS total
FROM LOAD(lineitem.csv)
UNION ALL
SELECT 'table' AS name, 10 AS price,
       0.05 AS discount, 9.5 AS total
\end{lstlisting}
Imperative-flavored declarative syntax has been repeatedly found to be more user-friendly than classic declarative syntax~\cite{Olston:2008:PLN:1376616.1376726}. 
Here however, it also serves to highlight the compositional nature of interactive views: each user action that changes the view's schema or contents is reflected in the script by a new statement appended to its end.  Thus, we aim for --- in principle at least --- a bi-directional mapping between user actions and statements in \langname.

In addition to enabling singletons and being easy to integrate with spreadsheets, the imperative flavor of \langname also enables backtracking and branching.  As illustrated in Figure~\ref{fig:hybridinterface}, users can quickly try hypothetical changes by checkpointing program state and applying new edits.  \sysname will support a comprehensive suite of branching and merging capabilities for both data~\cite{NA16} and workflows~\cite{SV08}.

\begin{figure}
\begin{lstlisting}[morekeywords={LOAD,REMOVE,REORDER,COLUMNS,SORT,ROW}]
s := UPDATE attribute = formula WHERE condition
   | ADD COLUMN colname
   | REMOVE COLUMN colname
   | INSERT ROW ( attribute = formula, ... )
   | DELETE WHERE condition
   | REORDER COLUMNS ( colname, colname, ... )
   | REORDER ROWS ( rowid, rowid, ... )
   | SORT ROWS sortorder
   
page := LOAD 'file' | page ; s
\end{lstlisting}
\caption{Grammar for \langname (\texttt{s} denotes a statement)}
\label{fig:grammar}
\vspace*{-6mm}
\end{figure}

\subsection{The Spreadsheet UI}


As a user edits tables and visualizations directly, these edits are reflected in the page where the table resides and are propagated to subsequent pages that depend on it. The user's edits, whether applied via the spreadsheet or notebook UI, are recorded as a form of workflow provenance~\cite{SV08,CF12a,AD11c,DC07}.  Our goal is not to reproduce the full interface of a spreadsheet, but rather to replicate as many of the
flexible data and schema manipulation features as possible within a more structured framework.  \sysname's UI allows users to:\\
\inlineitem{Overwrite arbitrary cells with constants, formulas, or regular expressions} Users may click on any cell in the output to overwrite its contents (as in a spreadsheet).  \\
\inlineitem{Cast cells to a new type} Dropdown menus allow the user to apply general transformations like typecasting.  The bulk transformation is applied to all cells in a selected region.  \\
\inlineitem{Copy/Paste cells} Users can copy and paste regions of cells.  The formula of the copied cell(s) is replicated in the target region through adapt\&apply.  If the target region is larger than the source, cells in the source region are tiled to scale over the entire target region.\\
\inlineitem{Add/Delete/Reorder columns or rows} Users may drag columns or rows to reposition them.  A tab at the bottom and right edges of the displayed table allows users to widen or lengthen the table, adding new columns or rows respectively.  Other interface elements allow users to insert rows (resp., columns) before or after any existing row (column).\\
\inlineitem{Sort data} A dropdown menu allows users to sort data according to values in one or more columns.\\
\inlineitem{Filter data} A dropdown menu allows users to filter out rows according to a formula defined over the row.\\
Many of these operations (e.g., paste, typecast) require the user to define a target, typically a rectangular area selected by clicking and dragging with the cursor; We also propose to support declarative regions, as discussed below.

\subsection{Spreadsheet to Notebook and Back}
To create a seamless interface between the spreadsheet and notebook UIs, we need to map operational semantics and effects between the two interaction models.  We now sketch solutions to several of the resulting challenges.

\tinysection{Identifying Singletons}
To allow singleton operations, \langname must be able to uniquely identify specific rows and columns of the dataset, including rows and columns introduced in the code itself.  More importantly, these identifying markers must persist through the program: A user edit applied to the row 10 of \texttt{'lineitem.csv'} must continue to be applied to the same entity, even if an insert operation occurs between rows 8 and 9.  We address this challenge through provenance: Each operation that creates rows generates a unique tag for each row, column, and cell, which persists through the lifetime of the row, column, or cell.

\tinysection{Positional vs Qualitative Semantics}
Spreadsheets allow formulas to reference other cells by relative position.  For example, a cell's formula might compute a cumulative average over all rows up to that point.  To capture these semantics, bulk update operations must permit a form of implicit windowing, semantics that can be unintuitive if handled incorrectly.  We address positional semantics as part of \langname's data model.

\tinysection{Readability}
Interactive spreadsheet interfaces encourage many small transformations.  In contrast, code promotes abstraction and terse expressions that precisely convey the user's intent.  As a result, directly translating visually generated operations into code is likely to produce a large, hard-to-follow, unreadable mess.  We address this by proposing a source-to-source readability-optimizing compiler below.


\tinysection{Formula Extraction}
An important challenge arises in the reverse direction as well.  When a user clicks on a formula to edit it, we need  to reconstruct the formula that derived the cell's value.  However, obtaining the precise formula may not be as simple as tracing the provenance of the cell's value, since operations (e.g., reordering rows) may alter dependencies.  We address this specific issue as part of our data model.


\section{The \langname Data Model}
\label{sec:language}

The fundamental unit of data in \langname is a \textit{cell}, a 3-tuple: $C = \tuple{id, f, v}$, consisting of a globally unique identifier $id$, a formula expression $f$, and a value $v$. The identifier of a cell is assigned to it when it is first allocated and is immutable --- even if the cell is moved to a different position in the spreadsheet.
By storing both a formula $f$ and its result $v$, a cell maintains data provenance akin to a provenance-aware data management system, where each record is associated with metadata describing how it was computed.  
Here, this metadata serves two purposes.  
First, as noted above, we need to be able to reliably materialize the formula backing each cell so that it can be edited.  We need to ensure that each operator defines precise semantics for how it affects formulas.
Second, and perhaps more importantly, we track both values and the formula used to derive them as a way to define operational semantics that minimize user surprise.  As we discuss shortly, one specific update to a spreadsheet may have many secondary, incidental effects on the spreadsheet's formulas and/or values.  By tracking both, we can better understand these effects and minimize the complexities and unexpected side-effects of each operation.

\tinysection{Coordinate System} Cells are arranged in a 2-dimensional grid of rows and columns indexed by a coordinate system, a function $s : \mathbb N \times \mathbb N \rightarrow id$ that maps positions in the grid to the cell occupying that position.  The function $s$ need not be complete, but must be one-to-one: a cell may only appear in one position in the spreadsheet.

\tinysection{Formulas} A formula is a primitive-valued expression that may include references to the values of other cells, identified by the cell's global id or by absolute coordinates (explicit and absolute references, respectively).  A formula evaluated in the context of a cell may also specify coordinate references as being relative to the cell (relative references).  Columns are usually denoted by letters and rows by numbers.
A \textit{state} is a 2-tuple $\tuple{ C, s }$ consisting of a set of cells $C = \{C_i\}$ and a coordinate system.  
We say that a formula $f$ evaluates to a value $v$ in the context of a given state ($f \mapsto_{\tuple{C,s}} v$) if, after replacing all references (coordinate references using $s$ and $C$, and explicit references using $C$), the formula reduces to $v$~\footnote{Similar operational semantics were previously proposed by Krishnamurthi and Ramakrishnan~\cite{Erwig2002}.}. 
We say that a state $\tuple{C, s}$ is \textit{valid}~\footnote{Note that this definition does not preclude direct or indirect circular references as long as the computations defined by the cell formulas have a fixpoint. However, such a fixpoint computation may be hard to understand for a user and, thus, we disallow circular references for now.} if each cell's formula evaluates to the cell's value:
$$\forall \tuple{id_i, f_i, v_i} \in C\;:\; f_i \mapsto_{\tuple{C,s}} v_i$$

User \textit{actions} in \langname transform a state $\tuple{C_1, s_1}$ into a new state $\tuple{C_2, s_2}$.   %
We call the semantics for an action correct if they ensure that if the input to an action is valid, then the action's output is also valid.

\begin{figure*}
\centering
{\small
\begin{subfigure}{0.3\textwidth}
  \centering
  \begin{tabular}{>{\tiny}rc|c|c}
   & \tiny A & \tiny B & \tiny C \\
    1& Alice & 10 & \texttt{=B1} (10)\\ \hline
    2& Bob & 4 & \texttt{=B2+C1} (14)\\ \hline
    3& Carol & 8 & \texttt{=B3+C2} (22)\\ \hline
    4& Dave & 9 & \texttt{=B4+C3} (31)
  \end{tabular}
  \caption{Initial State}
  \label{fig:rearrange:initial}
\end{subfigure}
\begin{subfigure}{0.3\textwidth}
  \centering
  \begin{tabular}{>{\tiny}rc|c|c}
   & \tiny A & \tiny B & \tiny C \\
    1& Alice & 10 & \texttt{=B1} (10)\\ \hline
    2& Carol & 8 & \texttt{=B2+C3} (22)\\ \hline
    3& Bob & 4 & \texttt{=B3+C1} (14)\\ \hline
    4& Dave & 9 & \texttt{=B4+C3} (31)
  \end{tabular}
  \caption{After swapping rows 2 and 3}
  \label{fig:rearrange:manual}
\end{subfigure}
\begin{subfigure}{0.3\textwidth}
  \centering
  \begin{tabular}{>{\tiny}rc|c|c}
   & \tiny A & \tiny B & \tiny C \\
    1& Alice & 10 & \texttt{=B1} (10)\\ \hline
    2& Dave & 9 & \texttt{=B2+C1} (19)\\ \hline
    3& Carol & 8 & \texttt{=B3+C2} (27)\\ \hline
    4& Bob & 4 & \texttt{=B4+C3} (31)
  \end{tabular}
  \caption{After sorting on column 'B'}
  \label{fig:rearrange:sort}
\end{subfigure}
}
\caption{Examples of both swapping rows and sorting rows in commercial database systems.}
\vspace*{-3mm}
\end{figure*}

\subsection{Unsurprising Inconsistencies}
User actions on a spreadsheet have both direct, intended effects, and may also have indirect, \textit{incidental} effects.  Examples include changing a formula (dependent formulas are recomputed), repositioning a row (formulas depending on the row are modified), or sorting (formulas are recomputed based on the new, sorted coordinate system).  
In commercial spreadsheet systems, indirect effect semantics can sometimes be inconsistent.  Take, for example, two mechanisms for rearranging rows in the table given in Figure~\ref{fig:rearrange:initial}.
A user might manually drag row 3 to a position between rows 1 and 2, effecting a swap of rows 2 and 3.  
Microsoft Excel, Apple's Numbers, and Google's Sheets~\footnote{These and other behaviors described were evaluated on Excel for Mac version 15.20, Numbers version 3.6.1, and Google Sheets as of April 2016.} all have identical behavior, each resulting in the table shown in Figure~\ref{fig:rearrange:manual}.  
Note that the formulas for C2 and C3 have changed to ensure that each cell retains its original value under the transposed coordinate system.  In other words, the user's \texttt{MOVE} action treats formula references as being \textit{explicit} references.  
Conversely, a user might sort the rows of the table in descending order on Column B.  The resulting table in all three systems is identical, and shown in Figure~\ref{fig:rearrange:sort}.  Here, the formulas in column C are changed only in appearance; each continues to reference the cells immediately to the left and above.  However the values of each cell have changed as a result.  In other words, the user's \texttt{SORT} action treats formula references as being \textit{relative} references.

Clearly, in spite of the superficial similarity between these two operations, their semantics are quite different.  However, viewed through the lens of \langname's data model, this design choice emerges as a form of the principle of least surprise~\cite{saltzer2009principles}.   Concretely, for purely structural operations (i.e., operations that simply manipulate the coordinate scheme), it is still necessary to propagate incidental effects to formulas and/or values.  The \texttt{MOVE} action translates cell formulas into the new coordinate scheme --- retaining stable values at the cost of changing formulas.  Meanwhile the \texttt{SORT} action re-evaluates cell-formulas under the new coordinate scheme --- retaining stable formulas at the cost of changing values.  By enforcing one of these two forms of stability (value or formula) for each user action, spreadsheet designers are guarding against hard to follow ``magic'' semantics.  We also note that \langname's data model admits both forms of stability.  

We tested a range of structural actions, and all consistently exhibited one of these types of stability: either on values (formulas are translated), or on formulas (new values are computed).  Our results are shown in Figure~\ref{fig:stability}.  Virtually all action semantics favor value stability --- clearly the simpler case in general.  Semantics that enforce formula stability are used primarily in sorting, which applies a non-intuitive, effectively random coordinate transform.  The other outlier is Numbers, where the cut operation removes data immediately, compared to Excel and Sheets, where the cut operation simply marks data to be moved on the subsequent paste.  This distinction allows Numbers to provide consistent paste semantics between cut and copy, while Excel and Sheets treat cut/paste as a special \texttt{MOVE}-like operation.  Specific tradeoffs aside, each system exhibits a preference for value-stability, falling back to formula-stability for non-intuitive coordinate transforms.

\begin{figure}
{\small
\begin{center}
\begin{tabular}{r|ccc}
 & & \textbf{Stability} & \\\hline
\textbf{Action} & \textbf{Excel} & \textbf{Numbers} & \textbf{Sheets} \\ \hline
Cut/Paste & V & F & V \\
Drag Cell/Row/Col & n/a & V & V\\
Insert Row/Col & V & V & V \\
Sort & F & F & F \\
Filter & V & V & V
\end{tabular}
\caption{Interface actions and whether they are \textbf{F}ormula-stable, or \textbf{V}alue-stable.  Excel does not support dragging.}
\label{fig:stability}
\end{center}
}
\vspace*{-7mm}
\end{figure}

\subsection{Regions to Relations}
Many operations in \sysname operate over sets or collections of cells.  
For example, aggregates in formulas, the `paste' operation, and type conversions all target or reference entire regions of cells. 
In a typical spreadsheet, such regions are specified as rectangular regions of cells in the current coordinate system (e.g., \texttt{[A3\;:\;B99]} or \texttt{[A\;:\;A]}).  
Conversely, in a relational setting, sets of target values are specified qualitatively through selection predicates.  

The former semantics are critical for enabling the spreadsheet interaction model, while the latter is important for generalizing the curation workflow beyond the initial dataset.  
Existing data curation systems focus on the latter approach; Even Wrangler~\cite{Kandel:2011:WIV:1978942.1979444}, which does allow users to initially write curation operators as singletons, still forces users to define a generalized predicate before moving on.

In \langname, regions combine both semantics. Concretely, a \textit{region} is defined through a 3-tuple $\tuple{X, Y, f}$, where $X$ is a (possibly infinite) set of columns, $Y$ is a (possibly infinite) set of rows, and $f$ is a boolean-valued formula defining a predicate over cells in the specified range. All cells in the intersection of $X$ and $Y$ fulfilling $f$ are part of the region.


\section{Workflow Rewriting}
\label{sec:generalizing}

As the user makes edits in the spreadsheet interface, the corresponding actions are recorded in the notebook as a \langname script.  
Although these scripts do encode the evaluation logic that generates the spreadsheet being displayed, they also serve as an audit trail, tool for reverting or altering older edits, and template for generalizing the same curation process to new data.
As such, \langname is subject to a different set of optimization goals than most programming languages.
Rather than optimizing for performance or resource usage as in a normal optimizer, \langname needs an optimizer that prioritizes both \textit{readability} and \textit{generality}.  

\subsection{Rewriting for Readability}
User actions on the spreadsheet are expected to be small, isolated changes.  
Recording them directly in this form is likely to produce long, hard to follow \langname scripts.  
Thus, it will be necessary for \sysname to dynamically rewrite scripts being modified in a principled way that optimizes for readability.
We consider readability to be a tradeoff between minimizing two measures: size and complexity.  
For example, consider a sequence of 10 update actions with the form:\\[-5mm]
\begin{lstlisting}[morekeywords={ROWID}]
UPDATE A = 3 WHERE ROWID = ?
\end{lstlisting}~\\[-5mm]
with \texttt{?} taking values from 1 to 10.  Instead, we could express all 10 updates in a single expression using a \texttt{BETWEEN} predicate that (a) more concisely represents the same concept, with (b) a similar level of complexity, and (c) is semantically equivalent.  
Similar transformations appear in optimizing compilers --- the above equivalence inverts a common compiler optimization called loop unrolling.  Although there has been substantial research effort on obfuscating compilers, we are not aware of any source-to-source compilers designed to increase code readability.
Vizier will not just record a \langname script for a workflow, but also keep track of what user operations each operation in the script is based on. This information can be exploited during rewriting. A set of formulas created by an adapt\&apply operation is a good candidate for rewriting, because we know that all formulas in such a set follow the same pattern.

\subsection{Generalizing Singletons}
Singletons allow users to try out hypotheticals, explore cleaning solutions, and conduct small-scale tests.  
It is often easier for users to perform one-off curation steps initially, repairing errors in the data as singletons, rather than expending the mental effort to generalize the repair upfront.
However, when the user needs to adapt their preliminary data cleaning solution to new data, to a larger dataset, or to an updated dataset, these singleton operations can become a burden.
Although they put more control over the curation process in the user's hands, singleton actions increase the size and complexity of a \langname script, with no benefits beyond the initial dataset.
In addition to considering readability-enhancing rewrites that preserve semantic equivalence, it will be necessary for \sysname to evaluate how singleton actions can be generalized --- effectively a form of query (or curation, in this case) by example~\cite{Zloof:1975:QE:1499949.1500034}.  Concretely, given a set of similar statements with singleton targets, we would like to propose to the user a set of rewrites that apply a single update to many (or all) of the singletons at once.  


\vspace{-.3cm}
\section{Related Work}
\label{sec:related}

Spreadsheet-style interfaces for relational data have been of interest to researchers and practitioners alike for some time now as a desperately needed form of direct data manipulation~\cite{JC07}.  Tyszkiewicz~\cite{Tyszkiewicz:2010:SRD:1807167.1807191} demonstrated an embedding of SQL into spreadsheet formula semantics. Excel provides database integration capabilities, and there is a spectrum of attempts at hybrid environments~\cite{Bakke:2011:SUI:1978942.1979313, Kandel:2011:WIV:1978942.1979444, Bendre:2015:DUD:2824032.2824121,Liu:2009:SAD:1546683.1547431,Witkowski:2005:QE:1083592.1083733}.

\tinysection{Spreadsheets with Workflows}
Trifacta/Wrangler~\cite{Kandel:2011:WIV:1978942.1979444} have features that are similar to \sysname.  As in \sysname, users generate curation workflows by directly editing data.  However, unlike \sysname, there is no support for singleton operations in the workflow language -- user edits must be generalized immediately through a recommendation interface.  
%
Query-by-Excel~\cite{Witkowski:2005:QE:1083592.1083733} (QBX) provides support for cube-style operations in a spreadsheet-like environment.  Although the goal is different, the mechanism is quite similar: QBX allows singleton outputs in the cube query, encoding them as \texttt{UPDATE} operations on the query output.  However, QBX treats only query outputs as mutable, while source data is fixed; \langname is free of this limitation.

\tinysection{Relational Spreadsheets}
SheetMusiq~\cite{Liu:2009:SAD:1546683.1547431} uses semantics for relational queries over spreadsheets.  Though superficially similar to \langname, it assumes static data, and does not attempt to preserve formula semantics through queries.
Related Worksheets~\cite{Bakke:2011:SUI:1978942.1979313} provide a spreadsheet UI for structured relational data, focusing on enabling strongly-typed data and foreign key references.  However, although editing cells is permitted, the work does not address cell formulas.
DataSpread~\cite{Bendre:2015:DUD:2824032.2824121} extends spreadsheets with relational database functionality: structured query support and a scalable relational engine for a backend.  By comparison, \langname starts with a structured relational data model and extends it with the illusion of freeform editing. 

\tinysection{Inspiration}
The idea of generalizing singleton operations is based on Query by Example~\cite{Zloof:1975:QE:1499949.1500034} and  Query by Explanation~\cite{Deutch:2016aa}.  As individual operations are grouped together, the system can learn to describe what the user is attempting to accomplish.  We plan to draw heavily on work in this area to develop \sysname's generalization engine.  
As the basis for the notebook-style interface and script provenance, we leverage work on scientific workflows~\cite{SV08,CF12a,DC07}, and we borrow ideas from 
reenactment~\cite{AG16} as the basis for \langname scripts.


\vspace{-.3cm}
\section{Conclusions}
\label{sec:conclusions}
We present our vision for Visier, a data curation system which exposes powerful curation operations through a UI that is a hybrid between the spreadsheet and notebook interface paradigms. In this work we focus on the user interface as well as present the initial design of a language \langname that can serve as the underlying computational model for operations in the system.  

\noindent\textbf{Acknowledgements: } \textit{
This work was supported in part by gifts from Oracle and NSF Grant CNS-1229185. Juliana Freire is partially supported by Defense Advanced Research Projects Agency (DARPA) MEMEX  program 
award FA8750-14-2-023.
Opinions, findings and conclusions 
expressed in this material are those of the authors and do not necessarily reflect the views of Oracle, the NSF, or DARPA.}


{\small
\bibliographystyle{plain}
\bibliography{HILDA}

\begin{thebibliography}{10}

\bibitem{AD11c}
Y.~Amsterdamer, S.~B. Davidson, D.~Deutch, T.~Milo, J.~Stoyanovich, and
  V.~Tannen.
\newblock Putting {Lipstick on Pig}: Enabling database-style workflow
  provenance.
\newblock {\em PVLDB}, 5(4):346--357, 2011.

\bibitem{AG16}
B.~Arab, D.~Gawlick, V.~Krishnaswamy, V.~Radhakrishnan, and B.~Glavic.
\newblock Formal foundations of reenactment and transaction provenance.
\newblock Technical report, IIT, 2016.

\bibitem{Bakke:2011:SUI:1978942.1979313}
E.~Bakke, D.~Karger, and R.~Miller.
\newblock A spreadsheet-based user interface for managing plural relationships
  in structured data.
\newblock In {\em SIGCHI}, 2011.

\bibitem{Bendre:2015:DUD:2824032.2824121}
Mangesh Bendre, Bofan Sun, Ding Zhang, Xinyan Zhou, Kevin Chen-Chuan Chang, and
  Aditya Parameswaran.
\newblock {DataSpread}: {U}nifying databases and spreadsheets.
\newblock {\em PVLDB}, 8(12):2000--2003, 2015.

\bibitem{Chan1996119}
Y.~E. Chan and V.~C. Storey.
\newblock The use of spreadsheets in organizations: Determinants and
  consequences.
\newblock {\em JIDM}, 31(3):119 -- 134, 1996.

\bibitem{CF12a}
F.~Chirigati and J.~Freire.
\newblock Towards integrating workflow and database provenance.
\newblock In {\em IPAW}, pages 11--23. 2012.

\bibitem{DC07}
S.~B. Davidson, S.~Cohen-Boulakia, A.~Eyal, B.~Lud\"{a}scher, T.~McPhillips,
  S.~Bowers, and J.~Freire.
\newblock {Provenance in Scientific Workflow Systems}.
\newblock {\em IEEE DEB}, 32(4):44--50, 2007.

\bibitem{Deutch:2016aa}
Daniel Deutch and Amir Gilad.
\newblock Learning queries from examples and their explanations.
\newblock {\em ArXiV}, 2016.

\bibitem{Erwig2002}
M.~Erwig and M.~Burnett.
\newblock {\em Practical Aspects of Declarative Languages}, chapter Adding
  Apples and Oranges, pages 173--191.
\newblock Springer, Berlin, Heidelberg, 2002.

\bibitem{JC07}
H.~V. Jagadish, A.~Chapman, A.~Elkiss, M.~Jayapandian, Y.~Li, A.~Nandi, and
  C.~Yu.
\newblock {Making database systems usable}.
\newblock {\em SIGMOD}, 2007.

\bibitem{Kandel:2011:WIV:1978942.1979444}
S.~Kandel, A.~Paepcke, J.~Hellerstein, and J.~Heer.
\newblock {Wrangler}: Interactive visual specification of data transformation
  scripts.
\newblock In {\em SIGCHI}, 2011.

\bibitem{Liu:2009:SAD:1546683.1547431}
B.~Liu and H.~V. Jagadish.
\newblock A spreadsheet algebra for a direct data manipulation query interface.
\newblock In {\em ICDE}, 2009.

\bibitem{NA16}
X.~Niu, B.~Arab, D.~Gawlick, O.~Kennedy Z.~H.~Liu, V.~Krishnaswamy, and
  B.~Glavic.
\newblock Provenance-aware versioned dataworkspaces.
\newblock In {\em TaPP}, 2016.

\bibitem{Olston:2008:PLN:1376616.1376726}
C.~Olston, B.~Reed, U.~Srivastava, R.~Kumar, and A.~Tomkins.
\newblock {Pig Latin}: {A} not-so-foreign language for data processing.
\newblock In {\em SIGMOD}, 2008.

\bibitem{saltzer2009principles}
J.~H. Saltzer and M.~F. Kaashoek.
\newblock {\em Principles of computer system design: an introduction}.
\newblock Morgan Kaufmann, 2009.

\bibitem{SV08}
C.~E. Scheidegger, H.~Vo, D.~Koop, J.~Freire, and C.~T. Silva.
\newblock {Querying and Re-using Workflows with VisTrails}.
\newblock In {\em SIGMOD}, 2008.

\bibitem{Tyszkiewicz:2010:SRD:1807167.1807191}
J.~Tyszkiewicz.
\newblock Spreadsheet as a relational database engine.
\newblock In {\em SIGMOD}, 2010.

\bibitem{Witkowski:2005:QE:1083592.1083733}
A.~Witkowski, S.~Bellamkonda, T.~Bozkaya, A.~Naimat, L.~Sheng, S.~Subramanian,
  and A.~Waingold.
\newblock {Query by Excel}.
\newblock In {\em VLDB}, 2005.

\bibitem{Yang:2015:LOA:2824032.2824055}
Ying Yang, Niccol\`{o} Meneghetti, Ronny Fehling, Zhen~Hua Liu, and Oliver
  Kennedy.
\newblock Lenses: An on-demand approach to etl.
\newblock {\em PVLDB}, 8(12):1578--1589, 2015.

\bibitem{Zloof:1975:QE:1499949.1500034}
M.~M. Zloof.
\newblock Query by example.
\newblock In {\em AFIPS}, 1975.

\end{thebibliography}
}
\end{document}